\documentstyle[twocolumn,epsf,aps,prb]{revtex}
\begin{document}
\title{Two Pseudogaps in the Cuprates}

\author{R.S. Markiewicz}
\address{Physics Department,
Northeastern U.,
Boston MA 02115}
\maketitle
\narrowtext

\begin{figure}
\leavevmode
   \epsfxsize=0.33\textwidth\epsfbox{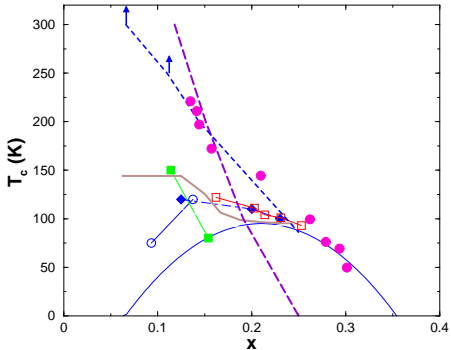}
\vskip0.5cm 
\caption{Doping dependence of T$_c$ (solid line) and superconducting 
fluctuations in Bi2212 (open circles) [\protect\onlinecite{Cor}] and (diamonds) 
[\protect\onlinecite{TIYM}], and YBCO (closed squares) 
[\protect\onlinecite{Bern}] and (open squares) [\protect\onlinecite{Mein}].  
Dotted line = leading-edge pseudogap from photoemission 
[\protect\onlinecite{Camp}] (arrows indicate temperatures are only lower 
limits); thick dashed line = weak pseudogap temperature $T^*$ in YBCO
[\protect\onlinecite{pseudT}]; closed circles = $\Delta /3$, where $\Delta$ =
peak position measured in tunneling [\protect\onlinecite{Miya}]; thick solid 
line = estimated superconducting temperature on a single stripe 
[\protect\onlinecite{MKI}].}
\label{fig:38a}
\end{figure}

Meingast, et al.\cite{Mein} measured the thermal expansion anomaly in 
underdoped YBa$_2$Cu$_3$O$_{7-\delta}$ (YBCO), interpreting their data as 
suggesting a connection between the pseudogap and preformed pairs.  In so doing,
they neglected early evidence\cite{StoP,EmBatl,TS} 
that most cuprates have {\it two} pseudogaps, and that only the lower of the 
two, the `strong' pseudogap, $T_{fl}$, is associated with superconducting 
fluctuations.  More recent evidence in both YBCO and Bi$_2$Sr$_2$CaCu$_2$O$_{8+
\delta}$ (Bi2212), Fig.~\ref{fig:38a}, confirms that the measured pseudogap 
temperatures fall into two groups.  The higher, `weak' pseudogap temperature 
$T^*$ is associated with a gap opening, with no clear evidence for 
superconducting fluctuations: it is measured from transport and heat capacity 
(dashed line)\cite{pseud,pseudT}, photoemission leading edge (dotted 
line)\cite{Camp}, and tunneling (filled circles = `peak' feature)\cite{Miya}.  
[The latter feature is associated with the total gap near $(\pi ,0)$; it is 
found that $2\Delta /k_BT^*\simeq 6$, consistent with a ratio 4.3 found by Ido, 
et al.\cite{Ido}.]  At a considerably lower temperature, $T_{fl}$, a clear onset
of strong superconducting fluctuations is found in underdoped samples.   Early 
evidence was based mainly on magnetic measurements; newer evidence includes 
magnetic measurements (Cu NMR $1/T_{2G}$ reduction)\cite{TIYM},
onset of Kosterlitz-Thouless fluctuations\cite{Cor}, and interlayer Josephson
tunneling\cite{Bern}.  The fluctuations found by Meingast, et al. clearly fall
into this latter group, extending to only about 2/3 of $T^*$ at the lowest
doping.  

Batlogg and Emery\cite{EmBatl} suggested that the weak pseudogap corresponds to 
the onset of electronic inhomogeneity (stripe fluctuations), the strong 
pseudogap to the onset of superconductivity on individual stripes, and the 
macroscopic superconducting transition $T_c$ is a signature of the
establishment of phase coherence between stripes.  It has also been 
postulated\cite{EmK} that superconducting pairing can be enhanced in the stripe
phase.  The data of Fig.~\ref{fig:38a} are very suggestive of this picture.  
First, the fact that there are no superconducting fluctuations near the weak 
pseudogap temperature $T^*$ makes it unlikely that $T^*$ is associated with 
preformed pairs.  Secondly, the strong pseudogap $T_{fl}$ is in good agreement
with a mean-field calculation\cite{MKI} (thick solid line) of the 
superconducting transition temperature on a charged stripe, assuming that 
$T_{fl}$ is proportional to the (pseudo-d-wave) gap along the stripe. 

Hence, the data of Meingast, et al., do not strongly support a model of 
preformed pairs, but are consistent with a stripe model for the pseudogap.

\end{document}